\documentclass[pra,amsmath,amssymb,twocolumn,showpacs,floatfix,nofootinbib]{revtex4}
\pdfoutput=1
\usepackage{graphicx}
\usepackage{dcolumn}
\usepackage{bm}

\begin{document}
\author{T. E. Wall}
\author{S. K. Tokunaga}
\author{E. A. Hinds}
\author{M. R. Tarbutt}\email{m.tarbutt@imperial.ac.uk}
\affiliation{Centre for Cold Matter, Blackett Laboratory,
Imperial College London, London SW7 2AZ, UK}

\title{Nonadiabatic transitions in a Stark decelerator}
\begin{abstract}

In a Stark decelerator, polar molecules are slowed down and focussed by an inhomogeneous electric field which switches between two configurations. For the decelerator to work, it is essential that the molecules follow the changing electric field adiabatically. When the decelerator switches from one configuration to the other, the electric field changes in magnitude and direction, and this can cause molecules to change state. In places where the field is weak, the rotation of the electric field vector during the switch may be too rapid for the molecules to maintain their orientation relative to the field. Molecules that are at these places when the field switches may be lost from the decelerator as they are transferred into states that are not focussed. We calculate the probability of nonadiabatic transitions as a function of position in the periodic decelerator structure and find that for the decelerated group of molecules the loss is typically small, while for the un-decelerated group of molecules the loss can be very high. This loss can be eliminated using a bias field to ensure that the electric field magnitude is always large enough. We demonstrate our findings by comparing the results of experiments and simulations for the Stark deceleration of LiH and CaF molecules. We present a simple method for calculating the transition probabilities which can easily be applied to other molecules of interest.
\end{abstract}


\maketitle

\section{Introduction}

Beams of fast-moving polar molecules can be brought to rest using a Stark decelerator \cite{Meerakker(1)09}. The molecules experience a force in the inhomogeneous electric field of the decelerator. By switching the field on and off in a carefully tailored sequence this force always opposes their forward motion. The Stark decelerator was first demonstrated for metastable CO \cite{Bethlem(1)99} and several other molecules have since been decelerated this way. The decelerator works best for molecules in low-field seeking states since they are easy to focus through the device. For the decelerator to work it is crucial that the molecules remain in the same quantum state throughout the deceleration process. The timing sequence that needs to be applied to the decelerator electrodes depends on the Stark shift and, since different states have different shifts, a molecule that changes state will no longer be decelerated. Moreover, if the state change is to one of the high-field seeking states, the molecule will be attracted to the electrodes where the field is strongest, and it is unlikely to be transmitted.

The unwanted change of state can occur because the molecules experience a time-varying electric field as they move through the decelerator and as the decelerator is switched from one field configuration to another. Transitions between states will only be significant if the time-varying electric field has frequency components that approach the frequency splitting between those states. Some simple order-of-magnitude reflections provide considerable comfort in this respect - in a typical deceleration experiment the Stark shifts are tens of GHz, whereas the molecule takes about 10\,$\mu$s to travel from one stage to the next and the rise and fall times ($\tau$) of the switching need only be 10 times smaller than that for the deceleration to be efficient. Nevertheless, we must consider carefully those cases where energy levels happen to be very close, as occurs at an avoided crossing, or when the electric field becomes so small that states differing only in their orientation with respect to the field are nearly degenerate. It is important to distinguish these two cases. In the decelerator, molecules need to traverse an avoided crossing {\it diabatically} so that the gradient of the Stark shift is the same on both sides of the crossing. That may be difficult to arrange if the energy gap at the avoided crossing is too large. Conversely, to preserve its orientation in low field, the molecule must follow the changing field {\it adiabatically}, which can be arranged by ensuring that the electric field in the decelerator is never too small.

When the decelerator switches, both the magnitude and the orientation of the electric field changes. It is the rotation of the electric field that is responsible for the orientation-changing transitions. On the axis of the decelerator the electric field vector rotates through 90$^{\circ}$ each time the decelerator switches. The rotation rate of the field vector depends on the position within the decelerator. At positions where the field magnitude is small prior to switching and large afterwards (or vice versa), the majority of the 90$^{\circ}$ rotation occurs in a small fraction of $\tau$, meaning that the rotation rate reaches values very much larger than $\tau^{-1}$. In fact, as we shall see, the rotation rate is at its largest at a point in space and time where the energy splittings are at their smallest and the molecules most vulnerable to nonadiabatic transitions. It is these rotation-driven transitions that are the main subject of this paper.

Losses due to nonadiabatic transitions and due to avoided crossings have been considered previously in the literature on Stark deceleration. For the deceleration of OH and NH in low-field seeking states, there is good experimental evidence that nonadiabatic transitions do not occur \cite{Meerakker(1)06, Hoekstra(1)07}. For the specific case of OH, this is supported by theoretical work \cite{Lara(1)08}. However, for LiH in the first rotationally excited state, numerical simulations could only be reconciled with experimental results by including in the simulations the probability of nonadiabatic transitions from the low-field to the high-field seeking component, driven by the rotation of the electric field during the switch \cite{Tokunaga(1)09}. In experiments on the deceleration of OH molecules in high-field seeking states, there was strong loss due to a crossing of hyperfine states which was eliminated by applying a small bias voltage so that the decelerator field was always above the field where the two states cross \cite{Wohlfart(1)08}. In experiments on SO$_{2}$, potential loss due to an avoided crossing of the decelerated state was considered, but occurred at too high a field to be important \cite{Bucicov(1)08}. Loss at an avoided crossing has been discussed in the context of decelerating asymmetric rotors \cite{Schwettmann(1)05}. Loss of molecules due to nonadiabtic transitions was also found to be significant in the recently demonstrated chip decelerator \cite{Meek(1)09}, and is important in traps for polar molecules, as observed in \cite{Kirtse(1)09} and discussed theoretically in \cite{Kajita(1)01, Kajita(1)06, Lara(1)08}.

We will first outline the theory of nonadiabatic transitions in a Stark decelerator. This leads to a simple adiabaticity criterion and to a simple method for calculating transition probabilities. We then provide some example calculations, showing that the transition probability is a strong function of position in the decelerator, and that molecules in the non-decelerated group are far more likely to be lost by this mechanism than those in the decelerated group. Finally, we compare the theoretical results with the results of experiments on the Stark deceleration of LiH and CaF molecules.

\section{Theory}\label{Sec:Theory}

\begin{figure}
\centering
\includegraphics[width=0.4\textwidth]{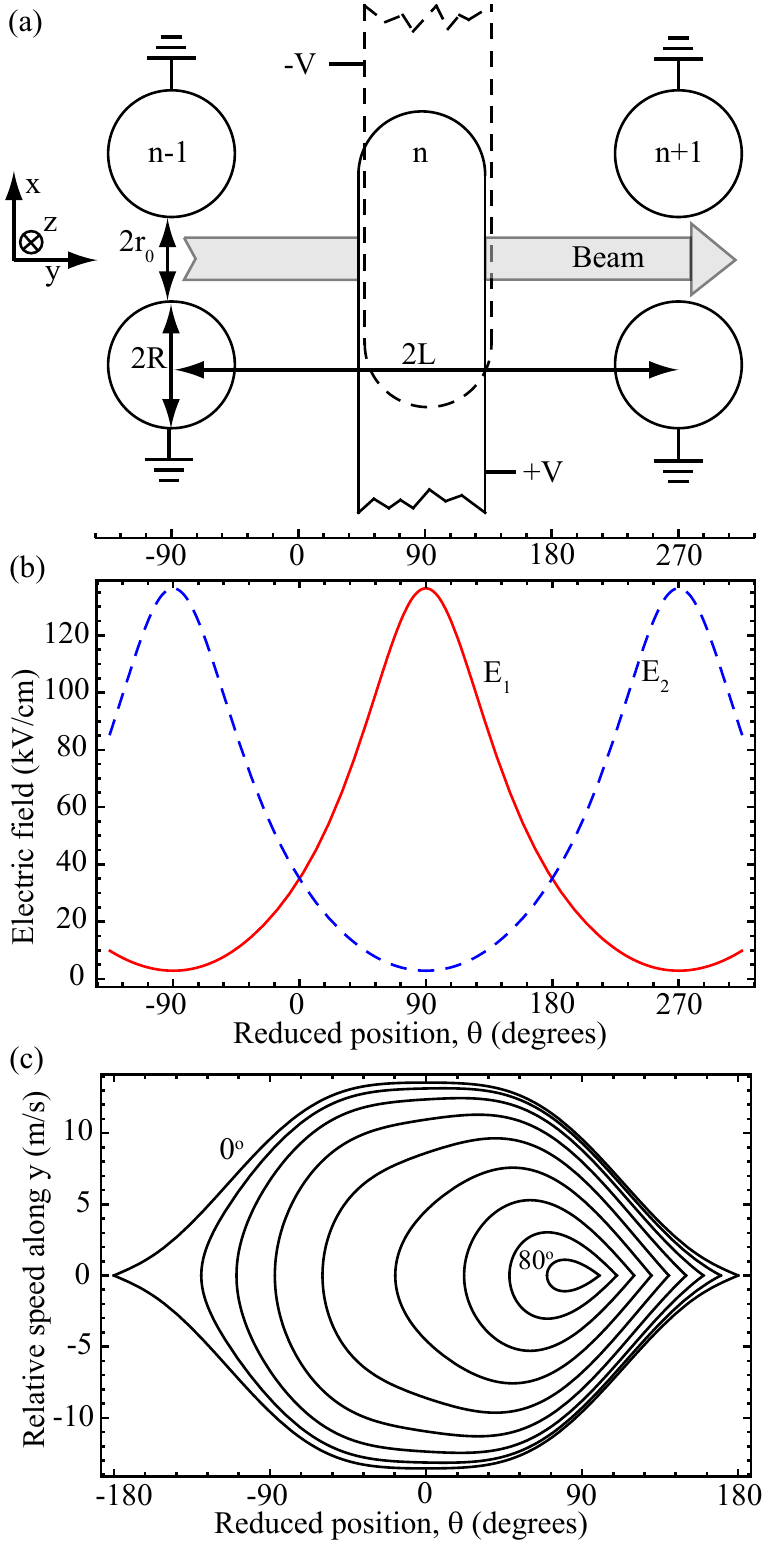}
\caption{(Color online) (a) Structure of the Stark decelerator, showing three consecutive deceleration stages centred on the $n$'th stage. Each stage is formed from a pair of rods whose axes are parallel to the $z$-axis for the odd-numbered stages, and to the $x$-axis for the even-numbered stages ($n$ is even). For clarity, the two rods forming the $n$'th stage are shown slightly displaced, though in reality one should be exactly behind the other. (b) Electric field profile versus reduced position for the two switch configurations when $R=1.5$\,mm, $r_{0}=1$\,mm, $L=6$\,mm, $V=15$\,kV. (c) Longitudinal phase-space acceptance for synchronous phase angles of $0^{\circ}$ (outermost line), through to $80^{\circ}$ (innermost line), in steps of $10^{\circ}$. The plot is for CaF molecules in the $|4,0\rangle$ state and for the electric field profile shown in (b).}
\label{Fig:decel}
\end{figure}

Figure \ref{Fig:decel}(a) illustrates the layout of a typical decelerator \cite{Bethlem(1)99}, showing three deceleration stages. Each stage is formed from a pair of parallel rods with the molecular beam passing between them. The axes of the rods are parallel to the $z$-axis in the odd-numbered stages and parallel to the $x$-axis in the even-numbered stages. The molecular beam travels along the $y$-axis and, to express the periodicity of the decelerator, the longitudinal coordinate is written (in radians) as the reduced position $\theta=\pi y/L$, $L$ being the distance between successive stages. The decelerator is switched between two states, state 1 which has the odd stages grounded and the even stages charged (as shown in Fig.\,\ref{Fig:decel}(a)), and state 2 which has the even stages grounded and the odd stages charged. The profile of electric field strength in these two states is shown in Fig.\,\ref{Fig:decel}(b) for a decelerator that has $R=1.5$\,mm, $r_{0}=1$\,mm, $L=6$\,mm and $V=15$\,kV. On the $y$-axis, the electric field is directed along $z$ in state 1 and along $x$ in state 2. When the decelerator switches between these two states the field on the beamline rotates by $90^{\circ}$ and changes in magnitude from the value given by the solid line to that given by the dashed line in Fig.\,\ref{Fig:decel}(b).

The switching sequence applied to the decelerator is synchronized to the motion of a molecule having a particular initial position and velocity. By construction, this synchronous molecule has the same value of $\theta$ (modulo $\pi$) every time the decelerator switches, and that value is called the synchronous phase angle, $\phi_{0}$. In the coordinates of Fig.\,\ref{Fig:decel}, $\phi_{0}=0$ corresponds to zero deceleration, and $\phi_{0}=90^{\circ}$ corresponds to maximum deceleration. The decelerator exhibits phase stability \cite{Bethlem(1)00}, meaning that it traps a group of molecules centred in phase space on the synchronous molecule. These phase-stable molecules remain close to the synchronous molecule throughout the deceleration process. The area they occupy in phase space is the longitudinal acceptance of the decelerator and depends on the degree of deceleration applied. Figure \ref{Fig:decel}(c) shows an example of this acceptance for several values of $\phi_{0}$. We will return to this figure to explain how the phase-stable molecules are less prone to nonadiabatic transitions than the non-phase-stable molecules.

We consider a rigidly rotating linear molecule of dipole moment $\mu$ and rotational constant $B$ in an electric field $\bf{E}$ which is $E_{1}\hat{\bf z}$ before the switch and $E_{2}\hat{\bf x}$ after the switch. During the switch, the first field decays with a time constant $\tau_{1}$ while the second field turns on with time constant $\tau_{2}$. The time-dependent Hamiltonian, $\hat{H}(t)$, that describes this system is given by
\begin{equation}
\hat{H}(t)=h B \pmb{\hat{N}}^2-\mu E_{1}e^{-t/\tau_{1}} \hat{C}_{0}^{(1)}-\frac{\mu E_{2}}{\sqrt{2}}(1-e^{-t/ \tau_{2}})(\hat{C}_{-1}^{(1)}-\hat{C}_{+1}^{(1)}),
\label{Eq:ham1}
\end{equation}
where $\pmb{\hat{N}}$ is the rotational angular momentum operator, and we have written the angular factors in the Hamiltonian in terms of the components of the rank 1 spherical harmonic operator, $\hat{C}^{(1)}_{q}=\sqrt{\frac{4\pi}{3}}\hat{Y}^{(1)}_{q}$.

\begin{figure}
\centering
\includegraphics[width=0.45\textwidth]{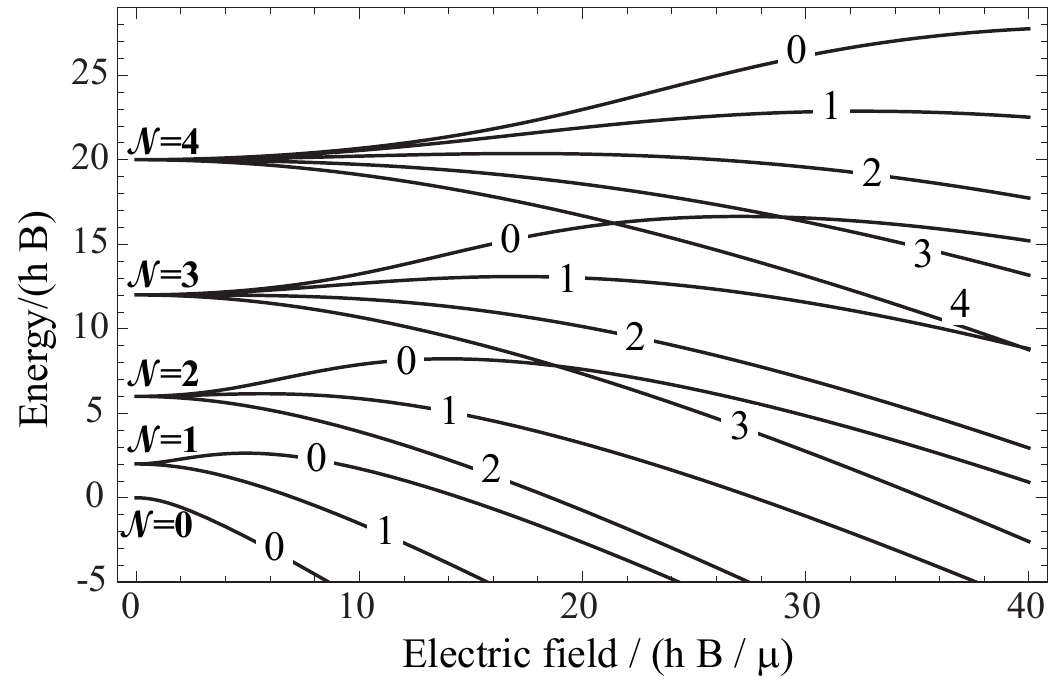}
\caption{Stark shift of the low-lying states of a rigid rotor molecule. States are labelled by their values of ${\cal N}$ and $M$.}
\label{Fig:Stark}
\end{figure}

In the absence of the electric field, the eigenstates are the angular momentum eigenstates $|N,M\rangle$, satisfying $\pmb{\hat{N}}^2 |N,M\rangle = N(N+1) |N,M\rangle$ and $\hat{N}_{z}|N,M\rangle=M|N,M\rangle$. The electric field mixes states of the same $M$ but different $N$ so that $N$ is no longer a constant of the motion.  We use the notation $|{\cal N},M\rangle_{E}$ to denote the mixed-$N$ eigenstates in the static field $E\hat{\bf z}$, omitting the subscript $E$ when it is not needed. To identify a specific state we assign a value to ${\cal N}$ which is equal to $N$ in the limit that $E\rightarrow 0$ adiabatically. These eigenstates, and the corresponding eigenvalues, $\epsilon_{{\cal N},M}$, are calculated by diagonalizing $\hat{H}(0)$ for many different values of $E_{1}$ using a suitably large basis set formed from the field-free eigenstates. Figure \ref{Fig:Stark} shows the energies of the lowest-lying states as a function of the electric field. Both the energy and the electric field are written in dimensionless units by dividing the former by the rotational constant $h B$, and the latter by the ratio of rotational constant to dipole moment $h B/\mu$. The Stark decelerator works best for the states with the largest positive Stark shift since these states are naturally focussed and are most strongly decelerated. Note that these low-field seeking states all turn over to become high-field seekers when the field is large, and that the turning point shifts to higher electric field for larger values of ${\cal N}$ and $B$.

At time $t=0$ the molecule is in the particular eigenstate $|{\cal N}_{p},M_{p}\rangle_{E_{1}}$. We can calculate the time evolution of the state vector $|\alpha\rangle$ in the $z$-fixed basis $|N M \rangle$, by solving the Schr\"{o}dinger equation for the coefficients $c_{N M}=\langle N M|\alpha\rangle$,
\begin{equation}
i \hbar \frac{d}{d t}c_{N M} = \sum_{N'M'}\langle N M|\hat{H}(t)|N'M'\rangle c_{N'M'},
\label{Eq:Schrodinger}
\end{equation}
with this initial condition and with $\hat{H}(t)$ given by Eq.\,(\ref{Eq:ham1}). Having found the time evolution of $|\alpha\rangle$, we can then calculate the probability of finding the molecule in any one of the states $|{\cal N},M\rangle_{E(t)}$ as a function of $t$.

Although this calculation is straightforward, the integration of Eq.\,(\ref{Eq:Schrodinger}) is slow because the basis set needs to include a large number of $N$ states and because the timescale of the switch is vastly larger than $1/B$, which sets the scale for the time step in the integration. Fortunately, a few simple transformations provide us with an equation that is much faster to solve and which also makes it obvious that it is the rotation of the electric field vector that drives the transitions.

If the coordinate system were rotated so that the field at time $t$ pointed along the local $z'$-axis, the instantaneous Hamiltonian would be $\hat{H}'=h B \pmb{\hat{N}}^2 - \mu E(t) \hat{C}_{0}^{(1)}$, $E(t)$ being the electric field magnitude at time $t$. Hamiltonian (\ref{Eq:ham1}) is related to this one by a rotation about the $y$-axis through an angle $\beta(t)$, $\hat{H}=\hat{D}^{-1} \hat{H}'\hat{D}$. Here $\hat{D}$ is the relevant rotation operator, $\hat{D}=\exp(-i \hat{N}_{y} \beta/\hbar)$. The evolution of the state vector $|\alpha\rangle$ is governed by the Schr\"{o}dinger equation: $i \hbar\,\frac{\partial}{\partial t}|\alpha\rangle = \hat{H}(t) |\alpha\rangle$. Applying the rotation operator to both sides of this equation, and introducing the state vector in the rotating frame, $|\alpha'\rangle = \hat{D} |\alpha\rangle$, we obtain
\begin{equation}
i \hbar\,\hat{D}\frac{\partial}{\partial t}(\hat{D}^{-1}|\alpha'\rangle) = \hat{D}\hat{H}|\alpha\rangle =
\hat{D} \hat{D}^{-1} \hat{H}' \hat{D}|\alpha\rangle = \hat{H}' |\alpha'\rangle.
\end{equation}
Using
\begin{equation}
\frac{\partial\hat{D}^{-1}}{\partial t}=\frac{i}{\hbar}\hat{N}_{y}\frac{d\beta}{d t}\hat{D}^{-1},
\end{equation}
and the fact that $\hat{D}^{-1}$ commutes with $\hat{N_{y}}$, we obtain the time evolution of the state in the rotating frame,
\begin{equation}
\left(\hat{H}'+\omega(t) \hat{N}_{y}\right)|\alpha'\rangle = i \hbar \frac{\partial}{\partial t}|\alpha'\rangle,
\label{Eq:SchrodingerRotating}
\end{equation}
where $\omega(t)=\frac{d \beta}{d t}$.

Next, following the algebra set out in App.\,\ref{AppA}, we express this equation in a basis formed from the instantaneous eigenvectors of $H'$, obtaining

\begin{equation}
\sum_{m}\left\langle m'\left|\omega(t) \hat{N}_y - \frac{i}{\omega_{m m'}} \frac{\partial \hat{H}'}{\partial t}\right|m\right\rangle a_{m}e^{-i \omega_{m m'}t} = i \hbar \frac{d a_{m'}}{dt},
\label{Eq:TimeEvolutionPretty}
\end{equation}
where $\hbar \omega_{m m'}=\epsilon_{m} - \epsilon_{m'}$, $a_{m}=\langle m|\alpha'\rangle e^{i \epsilon_{m} t/\hbar}$, $m$ stands for the pair of quantum numbers ${\cal N},M$, and $m'$ for ${\cal N}',M'$.

In Eq.\,(\ref{Eq:TimeEvolutionPretty}) the dynamics are expressed in a particularly transparent way. The first operator in the matrix element is due to the rotation of the electric field vector and couples states of different $M$, while the second arises from the change in the electric field magnitude and can only couple states of the same $M$ belonging to different manifolds ${\cal N}$. Provided the matrix elements in the sum vary little on the timescale of $1/\omega_{m m'}$, the terms in the sum oscillate around zero with an angular frequency $\omega_{m m'}$. Those amplitudes $a_{m}$ that are initially zero will always remain small unless they grow significantly in a time of order $1/\omega_{m m'}$. Thus, if the matrix elements are all much smaller than the energy splittings $\hbar \omega_{m m'}$, the state amplitudes hardly change meaning that the molecule adiabatically follows the changing field. This is the situation that we want in the Stark decelerator, so let us now estimate the transition probability arising from the off-diagonal terms. Assuming that these terms are at least small, we can treat them using time-dependent first-order perturbation theory. For a perturbation $\hat{V}$, the probability $P_{n}$ of being in state $|n\rangle$, having started out in the initial state $|i\rangle$, is at most $4|\langle n|\hat{V}|i\rangle|^{2}/(\hbar\omega_{ni})^{2}$ (see \cite{Sakurai} for example).

Taking first the last operator on the right hand side of Eq.\,(\ref{Eq:TimeEvolutionPretty}), we have $\langle n|\hat{V}|i\rangle=i\mu \frac{dE}{dt}\langle n|\hat{C}_{0}^{(1)}|i\rangle/\omega_{ni}.$ The maximum value of this last matrix element is 1, while that of $\frac{dE}{dt}$ is $E_{\rm max}/\tau$, $E_{\rm max}$ being the maximum electric field in the decelerator. As shown in Fig.\,\ref{Fig:Stark}, the weak-field seeking states have turning points, and this sets an upper limit to $E_{\rm max}$. Within manifold ${\cal N}$, the state with the largest positive Stark shift is $|{\cal N},M=0\rangle$, whose turning point is at a field of approximately $4\pi \hbar B {\cal N}^{2}/\mu$. For those states that are coupled by the perturbation, the smallest possible value of $\omega_{ni}$ occurs at low electric field and is $4\pi B {\cal N}$. Putting all this together, we conclude that the probability of a transition induced by the changing electric field magnitude is smaller than $1/(2\pi B \tau)^{2}$. Most molecules of interest have $B > 10$\,GHz, while the rise and fall time of the switch is typically $\tau > 100$\,ns, and so the maximum transition probability, about $10^{-8}$, is completely negligible.

Now we turn to the rotation of the field, i.e. to the matrix elements of $\omega \hat{N}_{y}$ in Eq.\,(\ref{Eq:TimeEvolutionPretty}). Here, the situation is quite different because this term couples states of different $M$ within the same manifold, and these states approach each other very closely when the field is small. The matrix element that we need is only non-zero when $M'=M \pm 1$, in which case it is
\begin{multline}
\langle {\cal N},M\pm1|\hat{N_{y}}|{\cal N},M\rangle = \mp \frac{i\hbar}{2}\sum_{N} \sqrt{(N\mp M)(N \pm M+1)} \times \\
\langle {\cal N},M \pm 1|N,M \pm 1\rangle \langle N,M|{\cal N},M\rangle.
\label{Eq:me1}
\end{multline}
At low electric field, where transitions are most likely to occur, the mixing of different $N$ states is small and so only one term ($N={\cal N}$) makes any significant contribution to the sum. The probability of finding the molecule in the state $M'=M\pm1$ is therefore
\begin{equation}
P_{\pm1}(t) < \frac{\omega^{2}}{\omega_{ni}^{2}} ({\cal N}\mp M)({\cal N} \pm M+1).
\label{Eq:perturbationResult}
\end{equation}
For low-lying rotational states, which are the ones of most interest to us, the transition probability will always be very small if $\omega \ll \omega_{ni}$ - this is the criterion for the rotation to be adiabatic.

\begin{figure}
\centering
\includegraphics[width=0.47\textwidth]{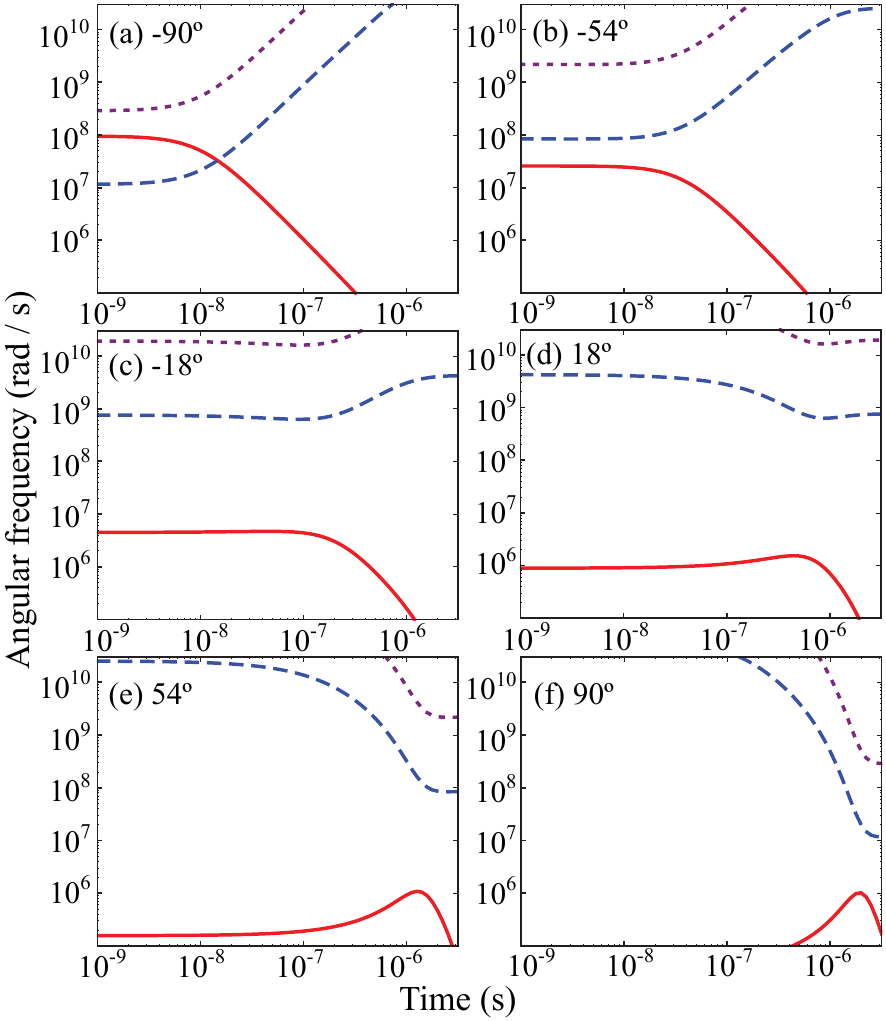}
\caption{(Color online) The rotation rate, $d\beta/dt$ (solid), the angular frequency separation of LiH $|1,0\rangle$ and $|1,1\rangle$ states (dotted), and the angular frequency separation of CaF $|4,0\rangle$ and $|4,1\rangle$ states (dashed), as a function of time since throwing the switch and at various reduced positions ($\theta$) within the decelerator. The switch rise and fall time was set to $\tau=500$\,ns, and the electric field profile used in the calculation is the one shown in Fig.\,\ref{Fig:decel}(b). Note the logarithmic scale on both axes.}
\label{Fig:Rates}
\end{figure}

\subsection{Examples}

We consider two examples, LiH prepared in the initial state $|i\rangle=|1,0\rangle$, and CaF prepared in the initial state $|i\rangle=|4,0\rangle$. They move through the decelerator illustrated in Fig.\,\ref{Fig:decel}(a), with $R=1.5$\,mm, $r_{0}=1$\,mm, $L=6$\,mm and $V=15$\,kV. We take $|n\rangle=|1,1\rangle$ for LiH and $|n\rangle=|4,1\rangle$ for CaF, and we take the rise and fall time of the switch to be $\tau=500$\,ns. Figure \ref{Fig:Rates} shows how the values of $\omega=d\beta/dt$ (solid lines), $\omega_{ni}^{\textrm{LiH}}$ (dotted lines) and  $\omega_{ni}^{\textrm{CaF}}$ (dashed lines) change as a function of time for several different values of the reduced position $\theta$. When $\theta=-90^{\circ}$, $\omega$ exceeds $\omega_{ni}^{\textrm{CaF}}$ for the first 15\,ns after the switch is thrown. At this position in the decelerator the field changes from a very small field along $z$ to a very large field along $x$ (see Fig.\,\ref{Fig:decel}(a,b)), and so the field vector completes almost all of the $90^{\circ}$ rotation in the first few ns, i.e. in a small fraction of $\tau$. As a result, the rotation rate is very large over this interval of time, $\omega \sim 10^{8}$\,rad\,s$^{-1}$. At the same time, the total field is still very small, and so the angular frequency interval between the two $M$ states is also very small, $\omega_{ni}^{\textrm{CaF}} \sim 10^{7}$\,rad\,s$^{-1}$. We see that the rotation is far from adiabatic. A CaF molecule that happens to be at this position when the field switches is very likely to end up in a different state. For LiH the situation is not as severe because, at early times, $\omega_{ni}^{\textrm{LiH}}$ exceeds $\omega$ by a factor of 3. Nevertheless, since the two frequencies are still of the same order of magnitude, we can expect there to be a significant transition probability. As we move away from this position in the decelerator, $\omega$ decreases as the difference in field magnitude between the two switch states decreases, and $\omega_{ni}$ increases because the minimum field value increases. At $\theta=-54^{\circ}$ for example, the maximum value of $\omega$ is about 3 times smaller than the minimum value of $\omega_{ni}^{\textrm{CaF}}$, while at $\theta=-18^{\circ}$ the former is more than two orders of magnitude smaller than the latter. At all positions, the Stark splitting for LiH is far greater than for CaF meaning that LiH is not as susceptible to nonadiabatic transitions. The situation at $\theta=90^{\circ}$ again seems dangerous - the field starts out very large along $z$ and ends up very small along $x$. This means that most of the rotation occurs at the very end of the switch time, at which time the frequency separation of the two states is very small. However, because the gradient of the exponential rise and fall is far smaller at late times than at early times, the rotation is not as fast and the maximum value of $\omega/\omega_{ni}^{\textrm{CaF}}$ is about 1/20.

\begin{figure}
\centering
\includegraphics[width=0.4\textwidth]{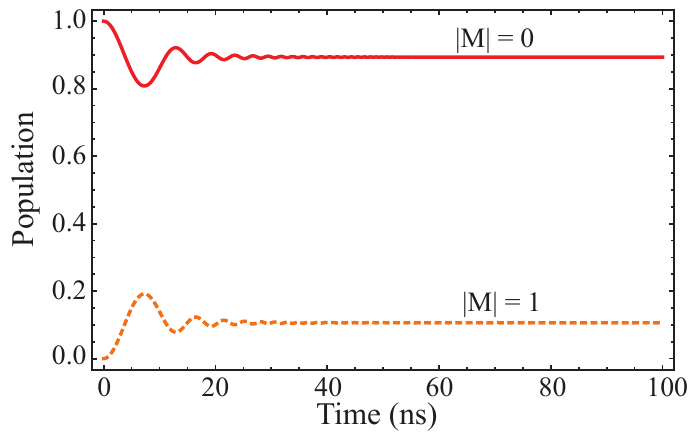}
\caption{(Color online) Evolution of the populations in the sublevels of the ${\cal N}=1$ manifold of LiH as a function of time since throwing the switch and at the position $\theta=-90^{\circ}$. The curves are labelled by their value of $|M|$. The initial state is $M=0$, the switch rise and fall time is $\tau=500$\,ns, and the electric field profile is the one shown in Fig.\,\ref{Fig:decel}(b).}
\label{Fig:LiHStateEvolution}
\end{figure}

To calculate the exact time-evolution of the state of the molecule, we set ${\cal N}'= {\cal N}$ in Eq.\,(\ref{Eq:TimeEvolutionPretty}), having already established that couplings between different ${\cal N}$ manifolds are negligible. Then we solve numerically the resulting set of differential equations,
\begin{equation}
\sum_{M=-{\cal N}}^{{\cal N}}\langle {\cal N},M'|\omega(t) \hat{N}_y |{\cal N},M\rangle a_{M}e^{-i \omega_{({\cal N},M)({\cal N},M')}t} = i \hbar \frac{d a_{M'}}{dt}.
\end{equation}

Figure \ref{Fig:LiHStateEvolution} shows what happens to LiH molecules initially in the $|1,0\rangle$ state, when they are at $\theta=-90^{\circ}$ at the moment of switching. Here, the rise and fall times of the switch are 500\,ns, and the electric field profile is as shown in Fig.\,\ref{Fig:decel}(b). As anticipated from the considerations above, the evolution is not fully adiabatic. The rapid rotation of the field vector that occurs during the first 10\,ns results in some population transfer into the $M=\pm 1$ states. After a few tens of ns the evolution becomes adiabatic and there is no further change in the state amplitudes. Moving away from this $\theta=-90^{\circ}$ position we find that the transition probability drops rapidly to zero. We note that in all cases the amplitudes of the $\pm M$ states are identical because the molecule starts in $M=0$ and, in our coordinate system, the $z$-axis lies in the plane of rotation. This symmetry is also seen by setting $M=0$ in Eqs.\,(\ref{Eq:me1}) and (\ref{Eq:perturbationResult}).

\begin{figure}
\centering
\includegraphics[width=0.4\textwidth]{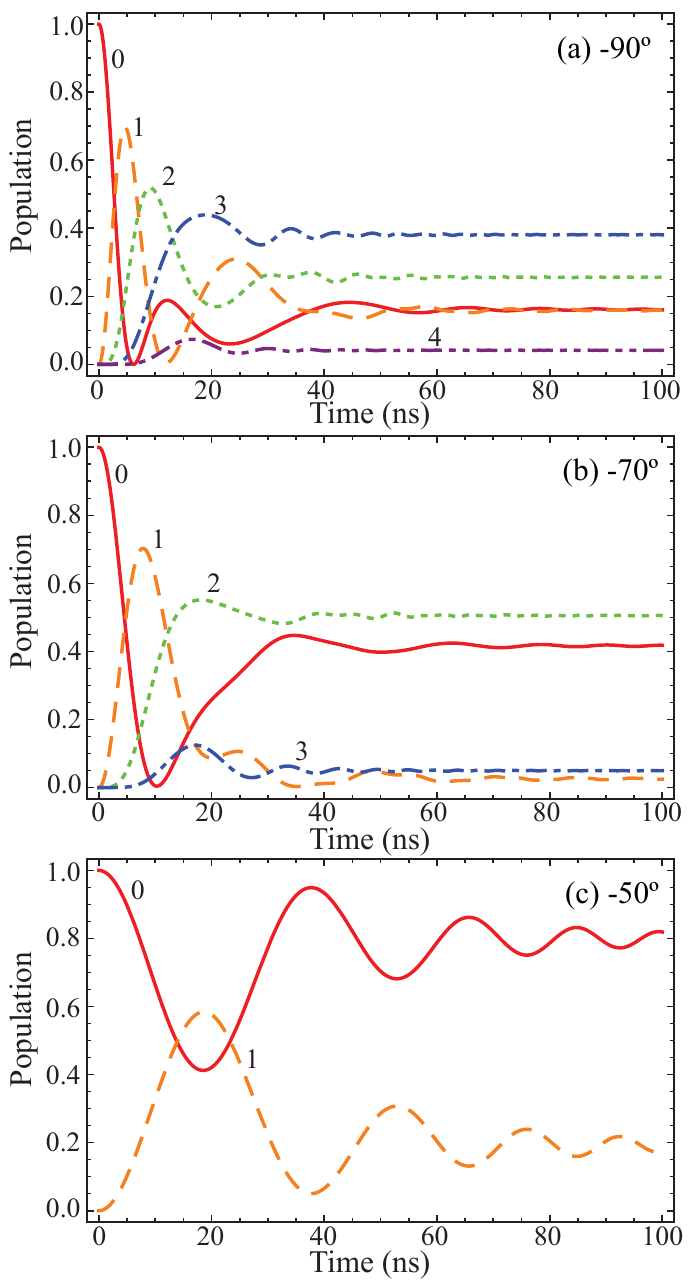}
\caption{(Color online) Evolution of the populations in the various sublevels of the ${\cal N}=4$ manifold of CaF as a function of time since throwing the switch, and at three different reduced positions in the decelerator. The curves are labelled by their value of $|M|$. The initial state is $M=0$, the switch rise and fall time is $\tau=500$\,ns, and the electric field profile is the one shown in Fig.\,\ref{Fig:decel}(b).}
\label{Fig:StateEvolution}
\end{figure}

Figure \ref{Fig:StateEvolution} shows what happens to CaF molecules initially in the $|4,0\rangle$ state, when they are at $\theta=-90^{\circ}$, $-70^{\circ}$ and $-50^{\circ}$. When $\theta=-90^{\circ}$, the initial population in the $|4,0\rangle$ state is rapidly transferred to the other $M$ states within ${\cal N}=4$. First, the $|M|=1$ state is populated, and from here the $|M|=2,3$ and 4 states all become populated as time progresses. After about 50\,ns the populations stop changing because the evolution changes over from being strongly nonadiabatic to being adiabatic, as anticipated from our consideration of Fig.\,\ref{Fig:Rates}(a). At $\theta=-70^{\circ}$ the evolution is slower and less of the population is driven to the higher $|M|$ states. In particular, the $|M|=3$ states, which received the largest share of the population at $\theta=-90^{\circ}$, now receive very little, and the $|M|=4$ states are never populated at all. These trends continue as $\theta$ increases, the evolution getting progressively slower and the population being confined more and more to the low $|M|$ states. At $\theta=-50^{\circ}$ all the population remains in the $M=0$ and $|M|=1$ states. Again, this is consistent with our expectations from Fig.\,\ref{Fig:Rates}(b) and Eq.\,(\ref{Eq:perturbationResult}). It is also interesting to study the situation at $\theta = 90^{\circ}$, where it is necessary to find the solution up to $t\sim3$\,$\mu$s so as to capture any dynamics that may occur when $\omega$ and $\omega_{ni}$ pass through their point of closest approach (see Fig.\,\ref{Fig:Rates}(f)). We find the population transfer to be negligible here.

\begin{figure}
\centering
\includegraphics[width=0.4\textwidth]{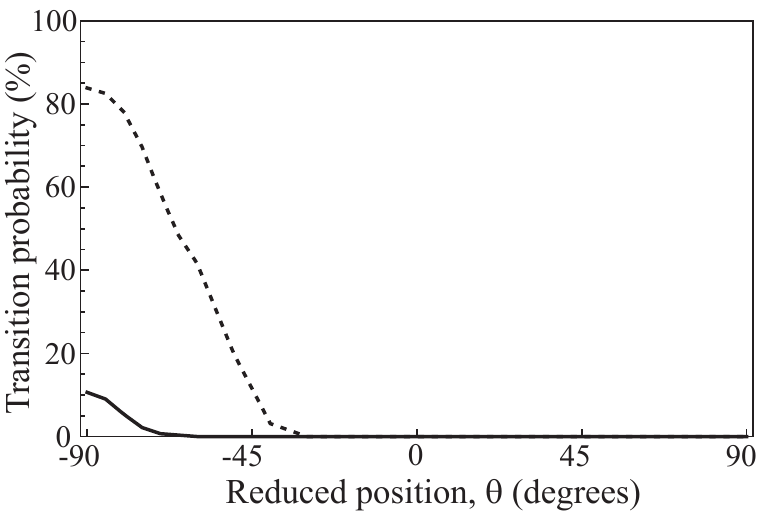}
\caption{Transition probability out of the $M=0$ state as a function of the reduced position in the decelerator. The position dependence is symmetric about $\theta=-90^{\circ}$. The switch rise and fall time is $\tau=500$\,ns, and the electric field profile is the one shown in Fig.\,\ref{Fig:decel}(b). Solid line: LiH in ${\cal N}=1$. Dashed line: CaF in ${\cal N}=4$.}
\label{Fig:TransitionProbs}
\end{figure}

Figure \ref{Fig:TransitionProbs} summarizes the results we have obtained. It shows the transition probability for LiH $|1,0\rangle$ and CaF $|4,0\rangle$ molecules as a function of the position in the decelerator at the moment of switching. Transitions occur in a zone centred on $\theta=-90^{\circ}$. For CaF the maximum transition probability is 84\% and the zone has a full width at half maximum of about $60^{\circ}$. For LiH, the maximum probability is 11\% and the zone has a full width at half maximum of about $20^{\circ}$. Obviously, faster switching or lower applied voltages will tend to increase the transition probabilities and extend the width of the zone.

\subsection{Loss}

In discussing the degree of molecule loss to be expected from these unwanted transitions, we need to distinguish molecules that are phase-stable from those that are not. The phase-stable molecules lie within the longitudinal phase-space acceptance of the decelerator; they are the ones being decelerated. The phase angle of these molecules, called $\phi$ and meaning the value of $\theta$ at the moment of switching, is constrained by the action of the decelerator to lie within the boundaries of the phase-space acceptance. These boundaries are shown in Fig.\,\ref{Fig:decel}(c) for various values of the synchronous phase angle, $\phi_{0}$. This figure shows that if the decelerator is operated at large phase angles, say $\phi_{0}>45^{\circ}$, corresponding to strong deceleration, the phase-stable molecules cannot reach the region centred on $\theta=-90^{\circ}$ where transitions are likely to occur. The group of molecules being decelerated are largely protected from transitions driven by the switching field. If we plot a molecule's phase-space position every time the decelerator switches we find that the phase-stable molecules move on closed trajectories centred on the synchronous molecule in the phase-space plot of Fig.\,\ref{Fig:decel}(c). Thus, even when the decelerator is operated at low phase angle, there will be a sizeable fraction of phase-stable molecules that never enter the regions where transitions are likely. By contrast, the molecules that are not phase-stable tend to explore all possible values of $\phi$ as they make their way through the decelerator. Since the region where transitions occur is sizeable, and since in a typical deceleration experiment the field switches about 100 times, we can expect a large fraction of those molecules to be driven into another state and be lost from the decelerator. So we reach the interesting conclusion that the loss will be far greater for the non-decelerated group of molecules than for the decelerated group.

\subsection{Hyperfine structure}

\begin{figure}
\centering
\includegraphics[width=0.47\textwidth]{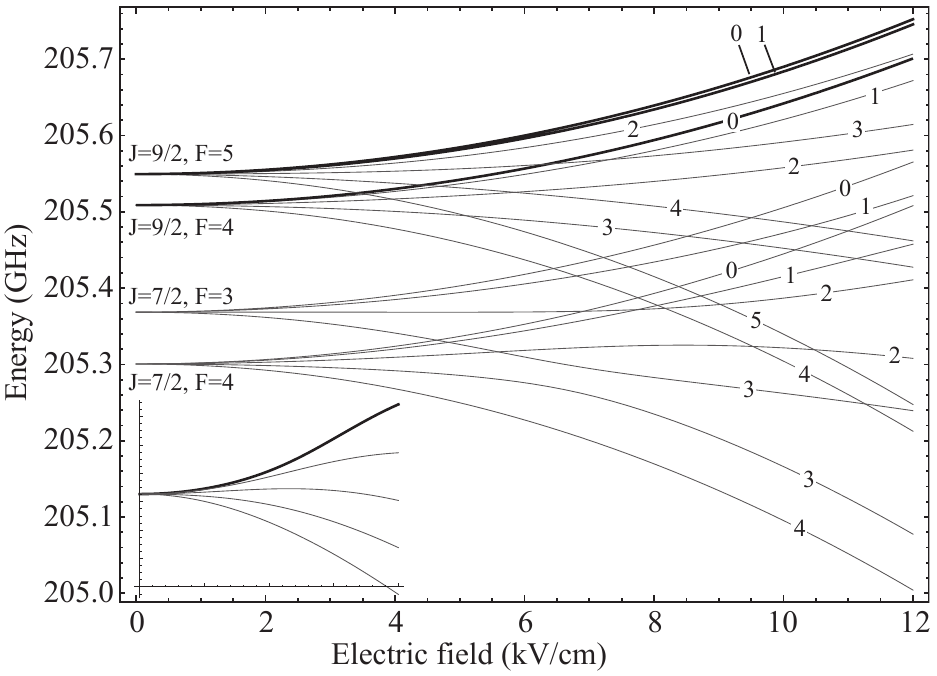}
\caption{The Stark shift of the energy levels of CaF that correlate to $N=4$ at low field. The states are identified by their values of $J$ and $F$ at low field, and by the value of $|M_{F}|$. At higher field the states separate into 5 groups that can be labelled by the projection quantum number $|M_{N}|$. Those that join the $M_{N}=0$ group are shown by bold black lines. The inset shows the same set of states up to a field of 200\,kV/cm.}
\label{Fig:Hyperfine}
\end{figure}

So far in our discussion, we have neglected hyperfine structure. This is acceptable when the hyperfine structure is small compared to the Stark shift at all relevant fields, as is the case for $^{7}$LiH in its $^{1}\Sigma$ electronic ground state. Here, the hyperfine splittings are of order 100\,kHz, and arise from the interaction of the electrons with the electric quadrupole moment of the Li nucleus, and from the interactions between the nuclear magnetic dipole moments and the magnetic dipole moment associated with the rotation. These splittings are small enough to be neglected in our calculations of nonadiabatic transitions. However, in CaF and many similar molecules, the hyperfine structure is comparable to the Stark shift at the fields strengths where we are troubled by nonadiabatic transitions. This complicates the matter quite considerably.

In CaF the main hyperfine interactions are a spin-rotation interaction, which splits a rotational state $N$ into two states labelled by $J=N \pm 1/2$, and the interaction between the electron spin and the spin of the fluorine nucleus, which further splits each state $J$ into two, labelled by the total angular momentum quantum number $F=J \pm 1/2$. Each state $F$ has $2F+1$ magnetic sublevels labelled by $M_{F}$, which is a good quantum number at all fields. In the absence of a magnetic field, states differing only by the sign of $M_{F}$ are degenerate. Figure \ref{Fig:Hyperfine} shows all the hyperfine states correlating to $N=4$ at low field, as a function of electric field. The main figure shows the Stark shifts at low fields, while the inset shows the same states up to a field of 200\,kV/cm. The energies were calculated by adding to $\hat{H}(0)$ (Eq.\,(\ref{Eq:ham1})) the usual hyperfine terms for a $^{2}\Sigma$ state, $\hat{H}_{\textrm{hyp}} = \gamma {\bf \hat{S}}\cdot{\bf \hat{N}} + b {\bf \hat{I}}\cdot{\bf \hat{S}} + c({\bf \hat{I}}\cdot{\bf k})({\bf \hat{S}}\cdot{\bf k}) + C {\bf \hat{I}}\cdot{\bf \hat{N}}$, where ${\bf \hat{S}}$ and ${\bf \hat{I}}$ are the electron and nuclear spin angular momentum operators, ${\bf k}$ is a unit vector along the internuclear axis, and $\gamma$, $b$, $c$ and $C$ are constants given for CaF in \cite{Childs(1)80}.

At very low field the energy levels group into 4 well separated components labelled by $J$ and $F$ because the Stark shift is small compared to the hyperfine structure. At high fields the energy levels regroup into 5 well separated components, labelled by the value of $|M_{N}|$, because the Stark shift is far greater than the hyperfine splitting. The tangle of levels at intermediate fields complicates our calculation of the transition probabilities. Instead of presenting this calculation, we make some general observations. First let us concentrate on a single $(J,F)$ manifold. It splits up into $M_{F}$ states in much the same way as do the $M_{N}$ states when we ignore the hyperfine structure (i.e. as in the inset). The level ordering is the same, and at any given field (over the range shown in the plot) the sizes of the splittings between the $M$ states are similar to those obtained in the absence of hyperfine structure. Since the matrix elements for the transitions between the $M$ states are also similar in the two cases we can conclude that the transition probabilities between $M_{F}$ states belonging to the same $(J,F)$ are not very different from the $M$-changing probabilites that we have already calculated. Continuing to focus on just one $(J,F)$ manifold, we note that a change of $M_{F}$ can change the character of the state from a low-field seeking to a high-field seeking one (compare the main figure with the inset). Again, this is no different from the situation with the hyperfine structure neglected. These considerations lead us to reason that loss due to nonadiabatic transitions is at least as severe when the hyperfine structure is included as when it is neglected. In addition to state changes within a $(J,F)$ manifold, transitions might occur between states from different manifolds, if these approach or cross one another. However, there is only a coupling if the values of $M_{F}$ differ by one. Focussing on the states with the largest positive Stark shift (bold black lines, all correlating to $M_{N}=0$ at high field), which are the ones of most interest to us, we see that such crossings or near approaches never occur. The same is true of the states that correlate to $|M_{N}|=1$. It is not true of the other states, but these are lost in the decelerator anyway. Putting these observations together, we conclude that our calculations of transition probabilities remain good estimates when the hyperfine structure is included. In most cases such rough estimates are sufficient since our task is usually to understand how severe the loss is likely to be and what can be done to avoid that loss.

\section{Experiment}

In this section, we discuss experiments on the Stark deceleration of $^7$LiH in the $|1,0\rangle$ state and of CaF in the $|4,0\rangle$ state, in order to demonstrate our main findings regarding nonadiabatic transitions.

\subsection{Setup}

The experimental setup for producing, decelerating and detecting LiH has been described before \cite{Tokunaga(1)07, Tokunaga(1)09}. We have not reported the deceleration of CaF previously, but the decelerator we use has the same structure as in \cite{Tokunaga(1)09} and the methods we use to produce and detect CaF are described in \cite{Wall(1)08}. Both molecular species are produced by laser ablation of a target into a supersonically expanding gas jet. For the LiH experiments, the target is Li and the carrier gas is a mixture of H$_2$ ($\approx 2\%$) and Kr. For the CaF experiments the target is Ca and the carrier gas is a mixture of SF$_6$ ($\approx 2\%$) and Xe. The temperature and pressure in the solenoid valve are approximately $300$\,K and 3--4\,bar respectively. The valve is fired at 10\,Hz and emits gas pulses with a duration of approximately 200\,$\mu$s into the source chamber which is maintained below $10^{-4}$\,mbar. The fundamental output of a Q-switched Nd:YAG laser, with a pulse duration of 10\,ns and an energy of 25\,mJ, is used to ablate the target outside the nozzle of the valve. The resulting beam passes through a 2\,mm diameter skimmer and into a second chamber which houses the decelerator and the detector. In the LiH source, most of the molecules are formed in the ground state and so we excite the molecules to the $N=1$ state upstream of the decelerator \cite{Tokunaga(1)09}. In the CaF source about 1\% of all the molecules are in the state of interest, the $|4,0\rangle$ state, and we do nothing to increase this fraction.

The Stark decelerator has 100 stages and has the structure shown in Fig.\,\ref{Fig:decel}(a), the dimensions being $r_{0}=1$\,mm, $R=1.5$\,mm and $L=6$\,mm. For the LiH experiments the centre of the first stage lies 135\,mm from the valve nozzle, while for the CaF experiments this distance was 126\,mm. Four high voltage switches are used to switch the electrodes between two voltages, $\pm V_{\textrm{hi}}$ and $\pm V_{\textrm{lo}}$. In Fig.\,\ref{Fig:decel}(a) $V_{\textrm{lo}}=0$, but in the experiments we can choose a different value in order to increase the minimum electric field in the machine, allowing the probability of nonadiabatic transitions to be controlled.

After passing through the decelerator the time-of-flight profile of the molecular pulse is recorded, with 10\,$\mu$s resolution, by cw laser induced fluorescence. The LiH molecules are detected 785\,mm from the nozzle by exciting the A$^{1}\Sigma^{+}(v'\!=\!4)$ -- X$^{1}\Sigma^{+}(v''\!=\!0)$ R(1) transition at 367.2\,nm. The CaF molecules are detected 810\,mm from the nozzle by exciting the A$^{2}\Pi_{1/2}(v'=0)$ -- X$^{2}\Sigma^{+}(v''=0)$ Q(4) transition at 606.3\,nm. The fluorescence is imaged onto the photocathode of a photomultiplier tube operated in photon counting mode. The detection is done at zero electric field and so cannot distinguish between the various $M$ sublevels. For CaF, the ground state hyperfine structure is resolved and for all data we detect on the $F''=5$ hyperfine component. Of the 11 sublevels of this component, 3 correlate to $M_{N}=0$ at high field, while the remaining 8 are distributed equally amongst the other 4 $M_{N}$ states (as shown in Fig.\,\ref{Fig:Hyperfine}).

\subsection{Results}

\begin{figure}
\centering
\includegraphics[width=0.4\textwidth]{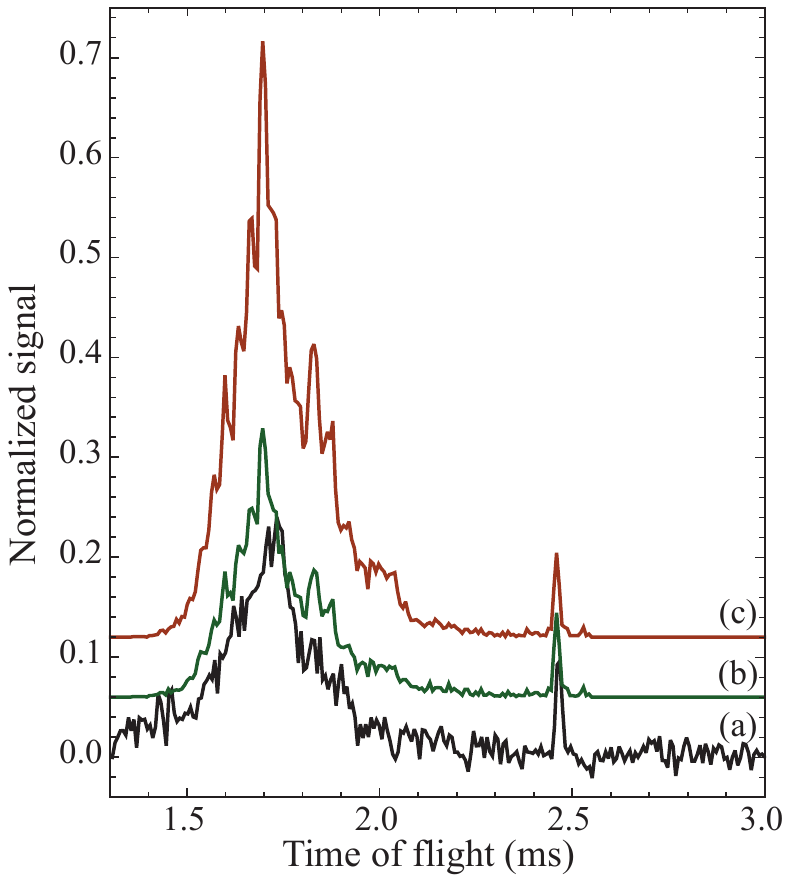}
\caption{(Color online) Time of flight profiles of LiH molecules decelerated from an initial speed of 420\,m/s to a final speed of 213\,m/s. The deceleration parameters are $V_{\textrm{hi}}=9.5$\,kV, $V_{\textrm{lo}}=0$, and $\phi_{0}=51^{\circ}$. (a) Lower trace, experimental data. (b) Middle trace, simulation including nonadiabatic loss. (c) Top trace, simulation for the adiabatic case. In all cases the signal is normalized to that obtained in dc mode. The traces are offset for clarity.}
\label{Fig:LiHResults}
\end{figure}

Figure \ref{Fig:LiHResults} compares experimental results for the deceleration of LiH with the results of simulations. In this experiment the electrodes are switched between $V_{\textrm{hi}}=9.5$\,kV and ground, and the decelerator is operated at a synchronous phase angle of $51^{\circ}$. The beam entering the decelerator has a mean speed of 437\,m/s and a temperature of 1.5\,K. The figure shows the LiH signal as a function of time, the zero of time being the moment when the ablation laser fires. The curves are normalized to the peak of the corresponding curves obtained when the voltage is applied constantly to the decelerator. The lower trace in the figure shows the experimental data, which was also discussed in \cite{Tokunaga(1)09}. The narrow peak that arrives at approximately 2.5\,ms corresponds to molecules decelerated from 420\,m/s to 213\,m/s, while the broad distribution centred near 1.7\,ms is due to the molecules that are outside the phase-space acceptance of the decelerator. The upper trace shows the result of a numerical simulation of the experiment where we assume that the molecules always follow the electric field adiabatically. The longitudinal velocity distribution used in the simulations is a Gaussian distribution with centre and width set equal to those measured in the experiments. The simulation does a good job of predicting the arrival time, width and amplitude of the decelerated bunch of molecules. The shape of the non-decelerated distribution is also reproduced satisfactorily, but its amplitude is not. Instead, the simulation predicts that this amplitude should be approximately 2.5 times larger than observed in the experiment.

Next, we modify the trajectory simulations to include the effects of nonadiabatic transitions in a simple way. The transition probability is first calculated as a function of the longitudinal position in the decelerator using the methods described in Sec.\,\ref{Sec:Theory}, for the voltage, rise time and fall time used in the experiment. We record the position of every molecule every time the decelerator switches, and then look up the probability of each molecule remaining in the $|1,0\rangle$ state after each switch. Multiplying together all these probabilities we obtain the total probability for each molecule to reach the end of the decelerator without switching state. We then take the detection probability to be proportional to this survival probability. The middle trace in Fig.\,\ref{Fig:LiHResults} shows the result of the modified simulation. As expected from the discussion above, we find that nonadiabatic transitions lead to loss of the non-decelerated part of the beam, but have no effect on the decelerated bunch of molecules. The modified simulation matches the experimental data very well indicating that this is the right loss mechanism, and that the simple procedure we have applied is a good one in this case. This procedure involves a number of simplifications. We take the transition probability to be independent of the transverse coordinates. We have neglected the hyperfine structure, which is a good approximation for LiH as discussed above. By simply multiplying transition probabilities, we are assuming that the superposition state rapidly decoheres, which is reasonable given the field inhomogeneity. Finally, instead of calculating the trajectories of molecules that have changed state we assume that they are lost rapidly from the decelerator by being pulled onto the electrodes. In principle, a LiH molecule transferred to the $|1,1\rangle$ state during one switch could be transferred back to the $|1,0\rangle$ state during a subsequent switch and so be recaptured in the decelerator before being lost. However, a LiH molecule in the $|1,1\rangle$ state is lost from the decelerator on the timescale of a few switch periods, and since the transition probability is quite small everywhere, there is very little chance of this recapture process occurring.

\begin{figure}
\centering
\includegraphics[width=0.4\textwidth]{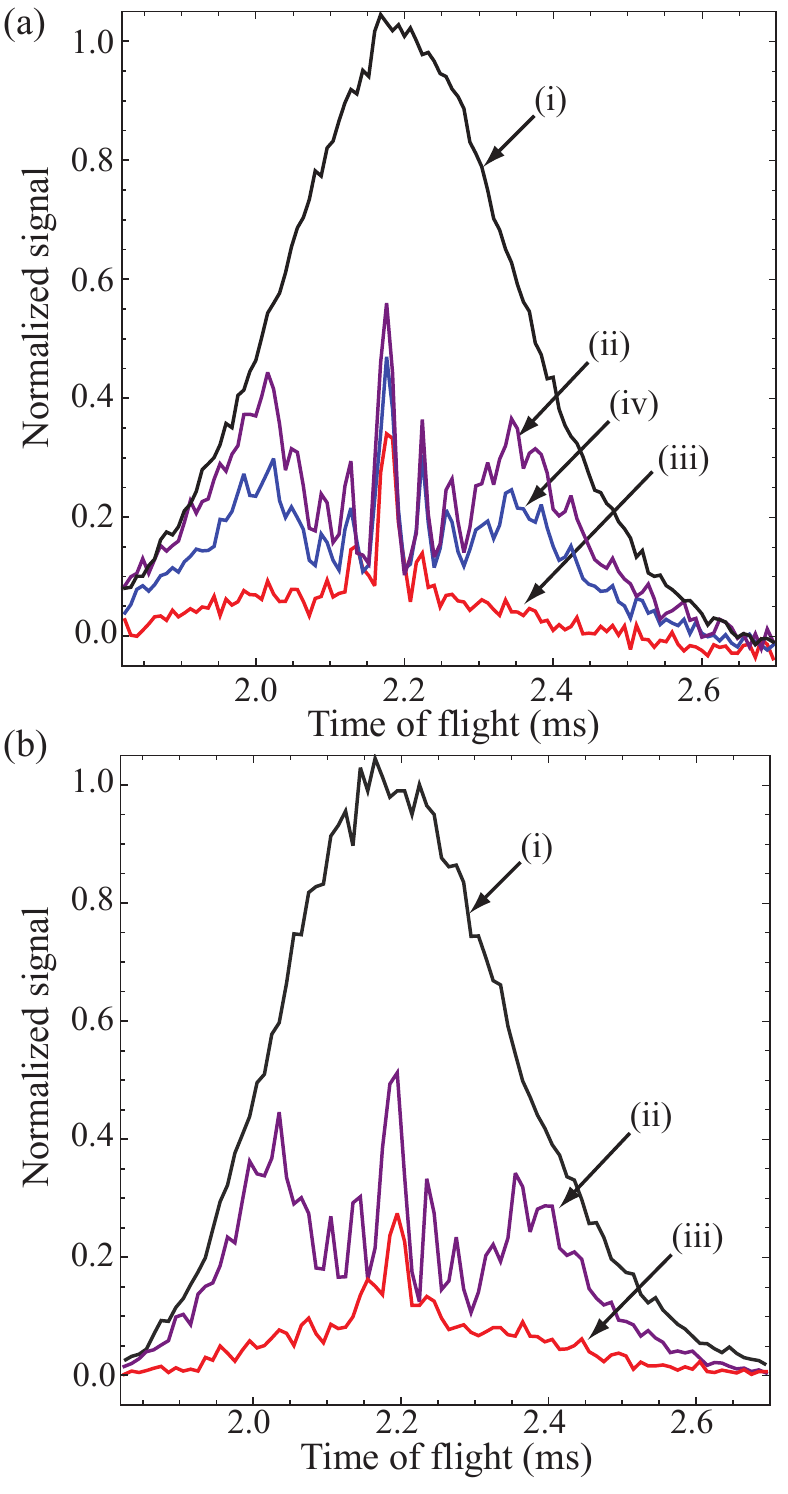}
\caption{(Color online) (a) Experimental time of flight profiles for CaF in the ${\cal N}=4$ state with a synchronous phase angle of $0^{\circ}$ (no deceleration), a synchronous speed of 370\,m/s, and an applied voltage of $V_{\textrm{hi}}=14$\,kV. (i) No switching. (ii) $V_{\textrm{lo}}=1.45$\,kV. (iii) $V_{\textrm{lo}}=0$. (iv) $V_{\textrm{lo}}=0.5$\,kV. (b) Simulations of these experiments. All the results share a common baseline.}
\label{Fig:CaFResults}
\end{figure}

Figure \ref{Fig:CaFResults}(a) shows experimental time-of-flight data obtained for CaF in the ${\cal N}=4$ state, when the decelerator high voltage is $V_{\textrm{hi}}=14$\,kV. Profile (i) is the one obtained when the high voltage is applied statically to all the electrodes of the decelerator. The molecules are focussed through the decelerator and the time-of-flight profile is a direct reflection of the velocity distribution in the beam. For this data, the mean speed is 368m/s and the translational temperature is 5\,K. The amplitudes of all the profiles are normalized to the peak of this one.

For the other three profiles in Fig.\,\ref{Fig:CaFResults}(a) the decelerator is switched with a synchronous phase angle of $0^{\circ}$ (no deceleration) and a synchronous speed of 370\,m/s. The only parameter that changes between these three profiles is the value of $V_{\textrm{lo}}$. For profile (ii) the decelerator was switched between $V_{\textrm{hi}}=14$\,kV and $V_{\textrm{lo}}=1.45$\,kV. In this case, the minimum electric field on the axis of the decelerator is raised to 12.0\,kV/cm, and the corresponding minimum splitting between the $M=0$ and $M=\pm 1$ states is $2.0\times 10^{8}$\,rad/s. This is larger then the maximum rotation rate (see Fig.\,\ref{Fig:Rates}) and we expect the evolution to be approximately adiabatic in this case. We observe a narrow central peak that corresponds to a bunch of molecules that are close in phase space to the synchronous molecule. This bunch is trapped in the moving potential well of the decelerator. We also see four smaller narrow peaks, two on either side of the main peak, and these correspond to bunches of molecules with speeds close to the synchronous speed but positions that are one and two potential wells ahead or behind the central bunch. Further out again are a pair of broader peaks corresponding to molecules that are not phase-stable, meaning that they are not trapped in the potential wells of the decelerator. Nevertheless, they are focussed through the decelerator, though not as effectively as in the case of static fields because the position-averaged field is only half as great when the decelerator is switching. We note that increasing $V_{\textrm{lo}}$ beyond this value of 1.45\,kV does not change the time-of-flight profile, indicating that the fully-adiabatic regime has been reached.  For profile (iii) the decelerator was switched between 14\,kV and ground, i.e. $V_{\textrm{lo}}=0$. For this case, the minimum field is 2.7\,kV/cm and the minimum splitting is $1.0\times 10^{7}$\,rad/s. The number of non phase-stable molecules is drastically reduced as we would expect since the maximum rotation rate is considerably larger than the minimum Stark splitting and so the loss due to nonadiabatic transitions is severe. The number of phase-stable molecules is also reduced but by a much smaller factor. This is what we expect for a synchronous phase angle of zero because, taking snapshots at each of the switch moments, the phase-stable molecules follow closed trajectories centred on $\theta = 0$ in the phase-space plot of Fig.\,\ref{Fig:decel}(c), and only some fraction of those trajectories overlap the zone around $-90^{\circ}$ where transitions are likely. The total number of molecules transmitted by the decelerator in this case is 22\% of the number transmitted in the adiabatic case. For profile (iv) the value of $V_{\textrm{lo}}$ is 500\,V. In this case, the minimum field is 5.3\,kV/cm and the minimum Stark splitting is $4.0\times 10^{7}$\,rad/s. This is still not sufficient to avoid transitions and we see that the degree of loss is intermediate between cases (ii) and (iii) as we would expect. Again, the broad wings of the time-of-flight profile are reduced in amplitude far more than the central narrow peak.

Figure \ref{Fig:CaFResults}(b) shows the results of simulating these CaF experiments. We simulate the motion of molecules in all 9 $M_{N}$ substates of $N=4$, and then use our knowledge of the correlation between the $M_{N}$ states at high field and the $F$ states at low field to calculate how many molecules are detected in $F=5$ (see Fig.\,\ref{Fig:Hyperfine}). Profile (i) is the result expected when the field is static, and serves to normalize the other profiles, exactly as for the experimental data. Almost all of the molecules transmitted are those in the $M_{N}=0$ and 1 states, as expected since these are the low-field seeking states. Profile (ii) is the result obtained when the decelerator is switching and we assume that molecules always follow the field adiabatically. Again, although we simulate all the $M_{N}$ states only the 0 and 1 states make a substantial contribution to the final signal. The result obtained agrees very well with profile (ii) of Fig.\,\ref{Fig:CaFResults}(a), showing that the molecules do indeed follow the field adiabatically once $V_{\textrm{lo}}=1.45$\,kV, and showing that their motion through the decelerator is well understood in this case. The central peak in the simulation is slightly broader and slightly smaller than in the experiments, but we find these features to be sensitive to the initial spatial distribution of molecules in the source, which is quite uncertain.

To obtain profile (iii) we have simulated the case of $V_{\textrm{lo}}=0$, where the loss due to nonadiabatic transitions is most severe. If we do this simulation in the same way as we did for the LiH case, assuming that molecules are immediately lost if they are switched out of the $M_{N}=0$ state, we find that the only molecules transmitted are the small bunch that are phase-stable. All of the un-trapped molecules are lost because the transition probability is high over a substantial region of the decelerator. This is not what happens in the experiments. Molecules are focussed in both the $M_{N}=0$ and $M_{N}=1$ states, and those that enter the decelerator in a high-field seeking state, or are switched into a high-field seeking state, may subsequently be switched to a low-field seeking state and thereby be transmitted. Indeed, the transition probabilities are so high that we can expect many of the un-trapped molecules to undergo several changes of state as they propagate through the machine. For the simulation that leads to profile (iii), we have tried to capture this complexity. We first calculate the position-dependent probability for transitions between each and every $M$ state, using the same parameters as in the experiment, neglecting the dependence of the probabilities on the transverse coordinates, and neglecting the hyperfine structure. For each molecule, we assign an initial $M$ state at random, solve the motion up to the moment of the first switch, re-assign the $M$ state according to the molecule's position and the pre-calculated probabilities, solve the motion up to the moment of the second switch, and continue with this procedure until the molecule crashes or exits the end of the decelerator. Unlike the adiabatic case, we find that molecules in all initial $M$ states contribute to the final signal, though 0 and $\pm 1$ still contribute the most. The simulated profile obtained this way agrees remarkably well with the experimental data, suggesting that the calculated transition probabilities are approximately correct despite the neglect of hyperfine structure, that the motion of the molecules in all relevant quantum states is well understood, and that the approximations we have made are reasonable.

\section{Conclusions}

In this paper we have shown, both theoretically and experimentally, that nonadiabatic transitions occurring in a Stark decelerator can result in a significant loss of molecules. When the decelerator switches, the rotation of the electric field vector drives transitions to other $M$ states, some of which are high-field seeking. The transition probability depends on how the rotation rate of the field vector compares to the splitting between neighbouring $M$ states. In order to ensure adiabatic behaviour, the electric field in the decelerator should always be large enough that the minimum splitting between states greatly exceeds the maximum rotation rate. This can be achieved by applying a bias electric field, and choosing switching rise and fall times that are as slow as possible while still being compatible with efficient deceleration. For a typical deceleration experiment, the maximum rotation rate is about $10^{8}$\,rad/s. This sets the scale for the size of the bias field required, which will depend on the molecular species. A magnetic field could also be used to suppress nonadiabatic transitions for some molecules, but its magnitude would need to be at least 0.01\,T for it to be effective. This would not be useful for $\Sigma$ states where the only relevant magnetic interaction is with the electron spin (if there is one). In this case it is the combination of the Zeeman interaction and the spin-rotation interaction that removes the degeneracy at zero electric field, and the induced splittings are small because the spin-rotation interaction is small.

We have shown that the transition probability is a strong function of the position in the decelerator, and that the decelerated group tends to be immune to nonadiabatic transitions because these molecules are always in a large field when the decelerator switches. This is particularly true when the synchronous phase angle is large. The fact that this loss mechanism distinguishes between the trapped and untrapped molecules could be used to filter out only those molecules that are being decelerated. This might be useful in experiments with slow molecules where the non-decelerated molecules present an unwanted background. Our experiments on the deceleration of LiH and CaF show clear evidence for loss due to nonadiabatic transitions. As expected, they show that the loss is far greater for the non-decelerated group of molecules and that this loss can be eliminated by applying a bias field of an appropriate size. The experimental results are in excellent agreement with simulations that include the transitions between the $M$ states.

We have presented a simple formalism for calculating the $M$-changing transition probabilities. The formalism is based on linear rigid rotor molecules, but with some modifications could also be applied to other molecules of interest, e.g. asymmetric rotors. Hyperfine structure complicates the calculation of the transition probabilities. However, if the hyperfine structure is small compared to the Stark shift at the minimum field encountered, as for LiH for example, it can be neglected. Our consideration of the hyperfine structure of CaF suggests that although this structure is large enough to be important, and though we have not done the complete calculations, the full results will not differ greatly from those obtained with hyperfine structure omitted.

For molecules that are not $\Sigma$ states, we expect $\Lambda$-doubling to suppress the probability of nonadiabatic transitions. Taking OH in the $^{2}\Pi_{3/2}, J=3/2$ state as an example, and neglecting hyperfine structure, $\Lambda$-doubling separates the low-field seeking components from the high-field seeking ones by almost $10^{10}$\,rad/s at zero field. Therefore, there are never any transitions between the two $\Lambda$-doubled states. Within the upper $\Lambda$-doubled manifold, the states differing in $M$ have very different Stark shifts, one being far larger than the other, and so transitions between these $M$ states, which are degenerate at zero field, could still result in loss. However, because of the relatively small splitting between the $\Lambda$-doubled components, the Stark shift becomes linear in relatively small fields. As a consequence, the Stark splitting of the two low-field seeking states is large enough to avoid nonadiabatic transitions for all electric fields typically found in the decelerator, even when there is no bias field. For example, this splitting is $2.6\times 10^{9}$\,rad/s at 2\,kV/cm.

Another interesting and relevant example is the $|J,K\rangle=|1,1\rangle$ state of ammonia. Its Stark shift changes from a quadratic to a linear dependence on the field magnitude once the Stark shift exceeds the inversion splitting. In $^{14}$NH$_{3}$ the inversion splitting is large and the Stark shift is quadratic for fields up to $\sim 20$\,kV/cm. At a field of 2\,kV/cm, the splitting between neighbouring $M$ states is $1.5\times 10^{8}$\,rad/s. In $^{14}$ND$_{3}$ on the other hand, the inversion splitting is far smaller and the Stark shift becomes linear for fields greater than $\sim 5$\,kV/cm. At a field of 2\,kV/cm, the splitting between neighbouring $M$ states is $1.9\times 10^{9}$\,rad/s, more than 10 times larger than for NH$_{3}$. These observations suggest that, unless a bias field is used, Stark deceleration experiments on NH$_3$ will exhibit loss of the non-decelerated group due to nonadiabatic transitions, whereas for ND$_3$ no such loss should occur.

\acknowledgements We thank Jony Hudson and Ben Sauer for helpful discussions regarding this work. We are grateful to Jon Dyne, Steve Maine and Valerijus Gerulis for their expert technical support. This work was funded by the UK EPSRC and STFC, and by the Royal Society. The research leading to these results has received funding from the European Community's Seventh Framework Programme FP7/2007-2013 under the grant agreement 216774.

\appendix
\section{Time evolution in an adiabatic basis}
\label{AppA}

Consider some general Hamiltonian which we write as a sum of two parts $\hat{H}(t) = \hat{H}_{1}(t) + \hat{H}_{2}(t)$, and where the instantaneous eigenvalues and eigenvectors of $\hat{H}_{1}(t)$ are $\epsilon_{m}(t)$ and $|m(t)\rangle$: $\hat{H}_{1}(t)|m(t)\rangle=\epsilon_{m}(t)|m(t)\rangle$. Using a basis formed by these instantaneous eigenvectors, we can write the Schr\"{o}dinger equation for the state vector $|\psi\rangle$ as
\begin{multline}
\sum_{m}\langle m'|\hat{H}_{1} + \hat{H}_{2}|m\rangle\langle m|\psi\rangle = i\hbar \sum_{m} \langle m'|\frac{\partial}{\partial t}(|m\rangle\langle m|\psi \rangle)\\
= i\hbar \sum_{m}\left(\langle m'|\frac{\partial}{\partial t}|m\rangle \langle m|\psi\rangle + \langle m'|m \rangle \frac{d}{dt}\langle m|\psi\rangle\right).
\end{multline}
Using the notation $c_{m}=\langle m|\psi \rangle$ we obtain the following equation for the time evolution of the coefficients $c_{m}$:
\begin{equation}
\sum_{m} \langle m' |\hat{H}_{1} + \hat{H}_{2}-i\hbar\frac{\partial}{\partial t}|m\rangle c_{m} = i \hbar \frac{d c_{m'}}{d t}.
\label{Eq:AppSchroForm1}
\end{equation}
Taking the derivative of $\hat{H}_{1}=\sum_{m''} \epsilon_{m''} |m''\rangle \langle m''|$ with respect to time we get
\begin{multline}
\frac{\partial \hat{H}_{1}}{\partial t} = \sum_{m''}\left[\left(\frac{d \epsilon_{m''}}{d t}\right) |m''\rangle \langle m''| + \right. \\
\left. \epsilon_{m''}\left(\frac{\partial|m''\rangle}{\partial t}\right)\langle m''| + \epsilon_{m''}|m''\rangle \left(\frac{\partial \langle m''|}{\partial t}\right)\right].
\end{multline}
It follows that for $m' \ne m$
\begin{multline}
\langle m'|\frac{\partial \hat{H}_{1}}{\partial t}|m \rangle = \epsilon_{m} \langle m'|\frac{\partial}{\partial t}|m\rangle + \epsilon_{m'}\left(\frac{\partial \langle m'|}{\partial t}\right)|m\rangle.
\end{multline}
The last term in this equation can be put into a more convenient form by noting that
\begin{equation}
\frac{\partial}{\partial t} (\langle m'|m \rangle) = 0 = \left(\frac{\partial \langle m'|}{\partial t}\right)|m\rangle + \langle m'|\frac{\partial}{\partial t}|m\rangle.
\end{equation}
Thus, for $m' \ne m$ we get
\begin{equation}
\langle m'|\frac{\partial \hat{H}_{1}}{\partial t}|m \rangle = (\epsilon_{m} - \epsilon_{m'})\langle m'|\frac{\partial}{\partial t}|m \rangle.
\end{equation}
Using this result in Eq.\,(\ref{Eq:AppSchroForm1}) we obtain
\begin{equation}
\epsilon_{m'}c_{m'} + \sum_{m} \left\langle m' \left|\hat{H}_{2}-\frac{i\hbar}{(\epsilon_{m}-\epsilon_{m'})}\frac{\partial\hat{H}_{1}}{\partial t}\right|m\right\rangle c_{m} = i \hbar \frac{d c_{m'}}{d t}.
\end{equation}
Finally, writing $c_{m}=a_{m}e^{-i\epsilon_{m} t/\hbar}$, and $\epsilon_{m}-\epsilon_{m'} = \hbar \omega_{m m'}$, we arrive at
\begin{equation}
\sum_{m} \left\langle m' \left|\hat{H}_{2}-\frac{i}{\omega_{m m'}}\frac{\partial\hat{H}_{1}}{\partial t}\right|m\right\rangle a_{m}e^{-i \omega_{m m'} t} = i \hbar \frac{d a_{m'}}{d t}.
\label{Eq:AppSchroForm2}
\end{equation}
This way of expressing the time evolution of the state vector makes it clear what operator is responsible for inducing transitions between the instantaneous eigenstates, which of the eigenstates are coupled by the time-varying Hamiltonian, and what condition must hold for the evolution to be adiabatic.

\end{document}